\documentclass[aps,prb,twocolumn,floatfix,showpacs,superscriptaddress]{revtex4}

\usepackage{graphicx}
\usepackage{bbm}
\usepackage{amsmath,amssymb}
\newcommand{\be}{\begin{equation}}
\newcommand{\beq}{\begin{equation}}
\newcommand{\ee}{\end{equation}}
\newcommand{\bea}{\begin{eqnarray}}
\newcommand{\eea}{\end{eqnarray}}
\newcommand{\ba}{\begin{array}}
\newcommand{\ea}{\end{array}}

\renewcommand{\vr} {{\bf r}}
\newcommand{\vj} {{\bf j}}
\newcommand{\vs} {{\bf s}}

\def\v#1{\mbox{\boldmath $#1$}}

\newcommand{\nn}{\nonumber}

\begin{document}
\title{Correlation energy of two-dimensional systems: \\
Toward non-empirical and universal modeling}
\author{S. Pittalis}
\email[Electronic address:\;]{pittalis@physik.fu-berlin.de}
\affiliation{Institut f{\"u}r Theoretische Physik,
Freie Universit{\"a}t Berlin, Arnimallee 14, D-14195 Berlin, Germany}
\affiliation{European Theoretical Spectroscopy Facility (ETSF)}
\author{E. R{\"a}s{\"a}nen}
\email[Electronic address:\;]{esa.rasanen@jyu.fi}
\affiliation{Institut f{\"u}r Theoretische Physik,
Freie Universit{\"a}t Berlin, Arnimallee 14, D-14195 Berlin, Germany}
\affiliation{European Theoretical Spectroscopy Facility (ETSF)}
\affiliation{Nanoscience Center, Department of Physics, 
University of Jyv{\"a}skyl{\"a}, FI-40014 Jyv{\"a}skyl{\"a}, Finland}
\author{C. R. Proetto}
\altaffiliation[Permanent addres: ]{Centro At{\'o}mico Bariloche and Instituto Balseiro, 8400
S.C. de Bariloche, R{\'i}o Negro, Argentina}
\affiliation{Institut f{\"u}r Theoretische Physik,
Freie Universit{\"a}t Berlin, Arnimallee 14, D-14195 Berlin, Germany}
\affiliation{European Theoretical Spectroscopy Facility (ETSF)}
\author{E. K. U. Gross}
\affiliation{Institut f{\"u}r Theoretische Physik,
Freie Universit{\"a}t Berlin, Arnimallee 14, D-14195 Berlin, Germany}
\affiliation{European Theoretical Spectroscopy Facility (ETSF)}

\date{\today}

\begin{abstract}
The capability of density-functional theory to deal with the ground-state of
strongly correlated low-dimensional systems, such as 
semiconductor quantum dots, depends
on the accuracy of functionals developed for the exchange
and correlation energies. 
Here we extend a successful approximation for the correlation 
energy of the three dimensional inhomogeneous electron gas, 
originally introduced by Becke [J. Chem. Phys. {\bf 88}, 1053 (1988)],
to the two-dimensional case. The approach aims to non-empirical 
modeling of the correlation-hole functions
satisfying a set of exact properties. Furthermore, 
the electron current and spin are explicitly taken into account. 
As a result, good performance is obtained in comparison with
numerically exact data for quantum dots with varying 
external magnetic field, and for the homogeneous two-dimensional 
electron gas, respectively.
\end{abstract}


\maketitle

\section{Introduction}
Density-functional theory~\cite{dft} (DFT) maps the 
complicated many-particle problem onto 
a simple one of non-interacting electrons
moving in an effective, local potential, the 
Kohn-Sham (KS) potential. The latter is constructed in
such a way that the ground-state density
of the non-interacting particles reproduces
the ground-state density of the interacting system.
Practical success of the approach 
depends on finding good approximations for the exchange-correlation (xc) energy
which, through functional derivation with respect to the
particle density, defines the xc part of the KS potential. 
Most of the approximations developed
so far have focused on three-dimensional (3D) systems, i.e.,
atoms, molecules, and solids. 
Many advances have been made beyond the commonly used local (spin) density 
approximation [L(S)DA] by means of, e.g., generalized gradient
approximations, orbital functionals, and hybrid functionals.~\cite{functionals}
Such efforts for two-dimensional (2D) systems have been relatively
scarce despite the rapidly increasing experimental and theoretical
interest in quasi-2D structures such as semiconductor layers and
surfaces, quantum-Hall systems, graphene, and various types of 
quantum dots~\cite{qd} (QDs).

When using DFT, QDs are most commonly
treated using the 2D-LSDA exchange~\cite{rajagopal} 
combined with the 2D-LSDA correlation parametrized first by 
Tanatar and Ceperley~\cite{tanatar} and later, with more 
satisfactory spin dependence, by Attaccalite {\em et al}.~\cite{attaccalite}
In many cases, the LSDA (prefix ``2D'' omitted below) 
performs relatively well compared, e.g., with
quantum Monte Carlo calculations.~\cite{henri} 
Nevertheless, there is a lack of 2D functionals to deal with
diverse few-electron QD systems, especially in the strong-correlation
regime. Only very recently, a local correlation functional was
developed in 2D within the Colle-Salvetti approach, which was
found to outperform the LSDA.~\cite{cs_paper} In its current
form, however, this local functional applies only to closed-shell
systems with zero spin and zero current.
In addition to the correlation-energy functional, 
new exchange-energy functionals have been developed for
finite 2D systems in our foregoing works.~\cite{becke_2D,becke_ring,GGA} 

In this paper we develop a correlation-energy 
functional in 2D. In the derivation, along the lines 
of the work of Becke~\cite{becke2} for 
the 3D case, we introduce a model for spin-dependent correlation-hole 
functions satisfying a set of exact properties in 2D.
As a result, we find a spin- and current-dependent approximation for the 
correlation energy. In comparison
with numerically exact results, the obtained accuracy is found to be
superior to the LSDA. 
However, we also find, and elucidate, that further
modeling of the dependency on the average electron density of those
parameters describing the size of the correlation-hole functions
in terms of the size of the exchange-hole (x-hole) functions would be required.
The applications of the functional to a set of few-electron QDs
with various relative amounts of correlation, ground-state
spins, electron currents, and external magnetic fields, 
confirm the overall usefulness of the approach.

\section{Theory}

Within the Kohn-Sham (KS) 
method of spin-DFT, \cite{BarthHedin:72} the ground-state energy and 
spin densities $\rho_{\uparrow}(\vr)$ and $\rho_{\downarrow}(\vr)$ 
of a system of $N=N_{\uparrow}+N_{\downarrow}$ interacting electrons are determined. 
The total energy of the interacting system is written as a 
functional of the spin densities~\cite{units}
\bea
E_{v}[\rho_{\uparrow},\rho_{\downarrow}] &=& T_s[\rho_{\uparrow},\rho_{\downarrow}] + 
\int{d\vr} \; v(\vr) \rho(\vr) \nn \\
&+& E_H[\rho] + E_{xc}[\rho_{\uparrow},\rho_{\downarrow}] 
\label{etot}
\eea
where $T_s[\rho_{\uparrow},\rho_{\downarrow}]$ is the kinetic energy functional 
of non-interacting electrons with spin densities $\rho_{\uparrow}$, 
$\rho_{\downarrow}$. 
$v$ is (at vanishing external magnetic field) the external (local) scalar potential acting 
upon the interacting system, $E_H[\rho]$ 
is the classical electrostatic or Hartree energy of the total charge density 
$\rho=\rho_{\uparrow}+\rho_{\downarrow}$, and  
$E_{xc}[\rho_{\uparrow},\rho_{\downarrow}]$ is the
xc energy functional. $E_{xc}$ may be further decomposed into the exchange energy, $E_x$,
and correlation energy $E_c$. We have already considered $E_x$ in
Ref.~\onlinecite{becke_2D}, and thus we here focus
on  $E_{c}$. Our starting point is the formal expression for $E_c$
in terms of the correlation-hole (c-hole) function,
\be\label{ECH}
E_{c}[\rho_{\uparrow},\rho_{\downarrow}]= \frac{1}{2} \sum_{\sigma\sigma'} \int d\vr_1 \int d\vr_2
\frac{\rho_{\sigma}(\vr_1)}{|\vr_1-\vr_2|}\,h^{\sigma\sigma'}_{c}(\vr_1,\vr_2),
\ee
where
\bea
h^{\sigma\sigma'}_{c}(\vr_1,\vr_2) \! \! \!  &=& \! \! \!
\int_{0}^{1} d\lambda \, h^{\sigma\sigma'}_{c,\lambda}(\vr_1,\vr_2) \nn \\ 
\! \! \! &=& \! \!\! \int_{0}^{1} d\lambda \, 
h^{\sigma\sigma'}_{\lambda}(\vr_1,\vr_2) -
h^{\sigma}_{x}(\vr_1,\vr_2)\delta_{\sigma\sigma'}
\eea
is the c-hole function. 
Here
$\{\sigma\sigma'\}=\{\uparrow\uparrow,\downarrow\downarrow,\uparrow\downarrow,\downarrow\uparrow\}$
and the parameter $\lambda\in[0,1]$
is the electronic coupling strength. \cite{dft}
In the above expression, the c-hole
corresponds to the {\em full} hole function $h^{\sigma\sigma'}_{\lambda}$
subtracted by the x-hole function defined as
\be
h^{\sigma}_{x}(\vr_1,\vr_2) =
-\frac{|\sum_{k=1}^{N_\sigma}\psi^*_{k,\sigma}(\vr_1)\psi_{k,\sigma}(\vr_2)|^2}
{\rho_{\sigma}(\vr_1)}.
\ee
Note that here $h^{\sigma}_{x}$ is defined to have a negative sign in
contrast to the standard definition [see, e.g., Eq. (4) in Ref.~\onlinecite{becke_2D}].
The $\lambda$-dependent hole function is given by
\be
h^{\sigma\sigma'}_{\lambda}(\vr_1,\vr_2) =
\frac{P^{\sigma\sigma'}_{2,\lambda}(\vr_1,\vr_2)}{\rho_\sigma(\vr_1)}-\rho_{\sigma'}(\vr_2),
\label{holefunction}
\ee
where $P^{\sigma\sigma'}_{2,\lambda}$ is the two-body reduced density
matrix (2BRDM) ~\cite{becke2}
\bea
P^{\sigma\sigma'}_{2,\lambda}(\v r_1,\vr_2) &=& N(N-1) \int d3 \int d4... \int dN \nn \\
&\times&\Psi_\lambda^*(1,2,...,N)\Psi_\lambda(1,2,...,N).
\label{2brdm}
\eea
Here, $\Psi_\lambda(1,2,...,N)$ stands for the ground-state many-body
wavefunction which is the exact solution of the electronic 
system with a Coulomb coupling strength $\lambda$,
$\int dN$ denotes the spatial integration and spin summation
over the $N$-th spatial spin coordinates $(\vr_N,\sigma_N)$, and we
have identified $\sigma_1=\sigma$, $\sigma_2=\sigma'$.
Note that the spin densities in Eq.~(\ref{holefunction}) are the same as
in the actual, fully interacting system.

As it is well known, 
in determining the correlation energy it is sufficient
to know the angular average of the c-hole function.
In 2D is natural to consider the  {\em cylindrical} average, given by
\be\label{DEFCA}
\bar{h}_{c,\lambda}^{\sigma\sigma'}(\vr,s) = \frac{1}{2\pi} \int_{0}^{2\pi} d \phi~ h^{\sigma\sigma'}_{c,\lambda}(\vr,\vr+\vs),
\ee
where $\vr_1=\vr$, $\vr_2=\vr+\vs$, and $\phi$ is the angle between $\vr$ and $\vs$.
The modeling of the c-hole functions is based on
satisfying the following conditions:
\begin{itemize} 
\item sum rule
\be\label{sumrule}
\int ds \,s\,\bar{h}^{\sigma\sigma'}_{c,\lambda}(\vr,s) = 0;
\ee
\item correct short-range behavior for $s \rightarrow 0$;
\item a proper decay in the limit $s \rightarrow \infty$;
\item characteristic size assumed to be proportional to
  the characteristic size of the corresponding angular averaged x-hole.
\end{itemize}

The short-range behavior of the c-hole can be worked out 
by considering the electronic cusp conditions for the 2D electronic wave function.~\cite{rajagopal2}
The angular average of the 2BRDM in Eq.~(\ref{2brdm}) is then given by 
\be
\bar{P}^{\sigma\sigma}_{2,\lambda}(\vr,s \rightarrow 0) 
\approx A_{\sigma\sigma}(\vr)\, s^2 \left( 1+ \frac{2}{3}\lambda s \right),
\ee
and
\be
\bar{P}^{\sigma\bar{\sigma}}_{2,\lambda}(\vr,s \rightarrow 0) 
\approx A_{\sigma\bar{\sigma}}(\vr) \left( 1+ 2\lambda s \right),
\ee
for the same-spin and opposite-spin elements, respectively. Note that 
the symbol $\bar{\sigma}$ always denotes spin opposite to $\sigma$, whereas
$\sigma'$, when used, can be equal to either $\sigma$ or $\bar{\sigma}$.
Here the coefficients $A_{\sigma\sigma'}(\vr)$ have a spatial dependence. 
Although the cusp condition as developed in
Ref.~\onlinecite{rajagopal2} pertains to the center-of-mass and 
relative coordinate, it can be shown, as noted by Becke,~\cite{becke2} 
that, at the order 
considered above, the same results apply also for the
coordinate system of $\vr$ and $\vs$.

The short-range behavior of the 2D x-hole is
known.~\cite{becke_2D,ELF} Thus, it is 
possible to consider the following model for the
same-spin and opposite-spin c-hole functions, respectively:
\bea
\bar{h}^{\sigma\sigma}_{c,\lambda}(\vr,s) &=&
\left[B_{\sigma\sigma}(\vr) -  D_{\sigma}(\vr) + 
\frac{2}{3} \lambda B_{\sigma\sigma}(\vr) s \right]s^2 \nn \\
& \times & F\left(\gamma_{\sigma\sigma}(\vr)\,s\right),
\label{h1}
\eea
and
\bea
\bar{h}^{\sigma\bar{\sigma}}_{c,\lambda}(\vr,s) &=&  \left[ 
B_{\sigma\bar{\sigma}}(\vr) - \rho_{\bar{\sigma}}(\vr)
+ 2 \lambda B_{\sigma\bar{\sigma}}(\vr)s \right] \nn \\
& \times & F\left(\gamma_{\sigma\bar{\sigma}}(\vr)\,s\right).
\label{h2}
\eea
Here $B_{\sigma\sigma}:= A_{\sigma\sigma}/\rho_{\sigma}$
and $B_{\sigma\bar{\sigma}}:= A_{\sigma\bar{\sigma}}/\rho_{\sigma}$
are coefficients to be determined, and
\be
D_{\sigma}:= \frac{1}{2} \left( \tau_\sigma - \frac{1}{4} \frac{\left( \nabla \rho_\sigma 
\right)^2}{\rho_\sigma} - \frac{\vj^2_{p,\sigma}}{\rho_\sigma} \right);
\label{D}
\ee
where 
$\tau_\sigma=\sum_{k=1}^{N_\sigma} |\nabla\psi_{k,\sigma}|^2$
is the (double of the) kinetic-energy density,
and
$\vj_{p,\sigma}=\frac{1}{2i}\sum_{k=1}^{N_\sigma} \left[
  \psi^*_{k,\sigma} \left(\nabla \psi_{k,\sigma}\right) - \left(\nabla \psi^*_{k,\sigma}\right) \psi_{k,\sigma} \right]$
is the spin-dependent paramagnetic current density.
In Eqs. (\ref{h1}) and (\ref{h2}),
the functions $F(\gamma_{\sigma\sigma'}(\vr)\,s)$ are introduced 
to ensure the decay of the c-holes
 in the limit $s \rightarrow \infty$. We choose them to have the form
$F(x)=\exp(-x^2)$, which seems appropriate in
the case of finite 2D systems. The parameters 
$\gamma_{\sigma\sigma'}(\vr)$ are determined by the zero-integral constraint of 
Eq.~(\ref{sumrule}) (see below).

Next, we introduce the characteristic sizes of the c-holes, 
$z_{\sigma\sigma}$ and $z_{\sigma\bar{\sigma}}$, for which the 
corresponding c-hole function vanishes, i.e.,
\be
\bar{h}^{\sigma\sigma}_{c,\lambda}(\vr,z_{\sigma\sigma})=0; \;\;\;\;
\bar{h}^{\sigma\bar{\sigma}}_{c,\lambda}(\vr,z_{\sigma\bar{\sigma}})=0.
\ee
Implying these conditions, the coefficients $B_{\sigma\sigma}$ and
$B_{\sigma\bar{\sigma}}$ in Eqs.~(\ref{h1}) and (\ref{h2}) can be written as
\be
B_{\sigma\sigma}=\frac{D_\sigma}{1+\frac{2}{3}\lambda
  z_{\sigma\sigma}}, \;\; 
B_{\sigma{\bar \sigma}} = \frac{\rho_{\bar{\sigma}}}{1+2\lambda
  z_{\sigma{\bar \sigma}}}.
\ee
It should be noted, however, that the size parameters
$z_{\sigma\sigma}$ and $z_{\sigma\bar{\sigma}}$
are functions of $\vr$. As for the 3D functional, we
assume the size of the c-hole to be proportional to the
size of the x-hole. Thus, we set
\bea
z_{\sigma\sigma}(\vr) & := & c_{\sigma\sigma} \left[
  |U_x^{\sigma}(\vr)|^{-1}+ |U_x^{\sigma}(\vr)|^{-1}\right] \nonumber
\\ & = & 2 c_{\sigma\sigma} |U_x^{\sigma}(\vr)|^{-1},
\label{cc1}
\eea
and
\be
z_{\sigma{\bar \sigma}}(\vr):= c_{\sigma{\bar \sigma}} \left[ |U_x^{\sigma}(\vr)|^{-1}+ |U_x^{{\bar \sigma}}(\vr)|^{-1}\right],
\label{cc2}
\ee
where $U_x^{\sigma}$ is the x-hole potential~\cite{becke_2D,becke}
for spin
$\sigma$, and $c_{\sigma\sigma}$ and $c_{\sigma{\bar \sigma}}$ are constants
to be determined (see below).
As suggested by Becke,~\cite{becke2} the proportionality is plausible
due to the fact that each electron is surrounded by its Fermi hole,
and hence the electrostatic interaction between two electrons can be expected 
to be screened beyond some characteristic length proportional to the average
size of the x-hole.
The argument is not spin-related, beyond
the fact that the characteristic length for the x-hole could be
different for spin up than for spin down. 
Of course, this is only
a simplification, but the physical picture is appealing, and
it has led to good results in 3D atomic systems.

The $\lambda$-dependent c-hole functions can now be written as
\be
\bar{h}^{\sigma\sigma}_{c,\lambda}(\vr,s)  = 
\frac{2}{3} \lambda D_{\sigma}(\vr) 
\left[ \frac{s-z_{\sigma\sigma}(\vr)}{1+ \frac{2}{3}  \lambda z_{\sigma\sigma}(\vr) } \right]
s^2\,F\left(\gamma_{\sigma\sigma}(\vr)\,s\right),
\ee
and
\be
\bar{h}^{\sigma{\bar \sigma}}_{c,\lambda}(\vr,s)  = 
2 \lambda \rho_{{\bar \sigma}}(\vr)
\left[ \frac{s-z_{\sigma{\bar \sigma}}(\vr)}{1+ 2  \lambda z_{\sigma{\bar \sigma}}(\vr) } \right]
F(\gamma_{\sigma{\bar \sigma}}(\vr)s) \; .
\ee
Integrating over $\lambda$ yields
\bea
\bar{h}^{\sigma\sigma}_c(\vr,s) & = & 
\frac{D_\sigma(\vr)\left[s-z_{\sigma\sigma}(\vr)\right]s^2F\left(\gamma_{\sigma\sigma}(\vr)s\right)}{2z_{\sigma\sigma}^2(\vr)}
\nonumber \\
& \times &
\left[2z_{\sigma\sigma}(\vr)-3\ln\left(\frac{2}{3}z_{\sigma\sigma}(\vr)+1\right)\right] , 
\eea
and
\bea
\bar{h}^{\sigma{\bar \sigma}}_c(\vr,s) & = &
\frac{\rho_{\bar \sigma}(\vr)\left[s-z_{\sigma{\bar \sigma}}(\vr)\right]F\left(\gamma_{\sigma{\bar \sigma}}(\vr)s\right)}{2z_{\sigma{\bar \sigma}}^2(\vr)}
\nonumber \\
& \times &
\left[2z_{\sigma{\bar \sigma}}(\vr)-\ln\left(2 z_{\sigma{\bar \sigma}}(\vr)+1\right)\right] \; . 
\eea

Finally, we enforce the sum rules in Eq.~(\ref{sumrule})
giving
\be
\gamma_{\sigma\sigma}(\vr) = \frac{3\sqrt{\pi}}{4 z_{\sigma\sigma}(\vr)},
\ee
and
\be
\gamma_{\sigma{\bar \sigma}}(\vr) = \frac{\sqrt{\pi}}{2 z_{\sigma{\bar \sigma}}(\vr)}.
\ee
This concludes the
derivation of the c-hole functions $\bar{h}^{\sigma\sigma}$ and
$\bar{h}^{\sigma{\bar \sigma}}$, apart from the determination of constants
$c_{\sigma\sigma}$, and $c_{\sigma{\bar \sigma}}$, respectively.

From the c-hole functions we can calculate 
the c-hole {\em potentials} as
\be
U^{\sigma\sigma'}_c(\vr)= 2 \pi \int_{0}^{\infty} ds \, h^{\sigma\sigma'}_{c}(\vr,s) \; .
\ee
For the same- and opposite-spin cases of our approximation, we find respectively
\bea
U^{\sigma\sigma}_c(\vr) & = & \frac{16}{81\pi}\left(8-3\pi\right) D_\sigma(\vr)
z^2_{\sigma\sigma}(\vr) \nonumber \\
& \times & \left[2 z_{\sigma\sigma}(\vr)-3\ln \left( \frac{2}{3}z_{\sigma\sigma}(\vr)+1\right)\right],
\eea
and 
\bea
U^{\sigma{\bar \sigma}}_c(\vr) & = & (2-\pi)\rho_{{\bar \sigma}}(\vr) \nonumber \\
& \times & \left[2
  z_{\sigma{\bar \sigma}}(\vr)-\ln\left(2 z_{\sigma{\bar \sigma}}(\vr) +1\right)  \right].
\eea

The correlation energies are given by
\be
E^{\sigma\sigma'}_{c} = \frac{1}{2} \int d\vr\, \rho_{\sigma}(\vr)
\,U^{\sigma\sigma'}_{c}(\vr) \; .
\ee
Thus
\be
E_c [\rho_{\uparrow},\rho_{\downarrow}] =E^{\uparrow\uparrow}_{c} + E^{\downarrow\downarrow}_{c} + 
2 E^{\uparrow\downarrow}_{c},
\ee
where we have used the condition $E^{\uparrow\downarrow}_{c}=E^{\downarrow\uparrow}_{c}$.

Alternatively, we can compute the correlation energy directly from
the c-hole functions. From Eqs. (\ref{ECH}) and (\ref{DEFCA})
we get
\be
E_{c}[\rho_{\uparrow},\rho_{\downarrow}]= \pi \sum_{\sigma\sigma'} \int d\vr
 \int ds ~
\rho_{\sigma}(\vr)\,\bar{h}^{\sigma\sigma'}_{c}(\vr,s).
\ee

We remind that $D_\sigma(\vr)$ 
introduced in Eq.~(\ref{D}) vanishes for all the single-particle
($N=1$) systems.\cite{becke2}
Therefore, the c-hole and thus $E_c$ vanish as well.
In other words, our approximation for the correlation energy is 
self-interaction free for $N=1$.
~\footnote{This is due to the fact
that for $N=1$ systems, 
$\psi_\sigma(\vr)=\exp[i\theta(\vr)]\sqrt{\rho_\sigma(\vr)}$, without loss of generality.
Evaluation of the three contributions to
$D_\sigma(\vr)$ in Eq.~(\ref{D}), for this particular case, yields
$D_\sigma(\vr)=0$} 

\section{Numerical Results}

The first task in the numerical applications is to 
complete the correlation functional by finding
approximations for constants $c_{\sigma\sigma}$ and 
$c_{\sigma{\bar \sigma}}$ in Eqs.~(\ref{cc1}) and (\ref{cc2}).
For this purpose, we consider a set of harmonically
confined QDs, where the external confinement is given by
$v(r)=\omega^2 r^2/2$. Reference results for the correlation
energies can be obtained from 
\be
E^{\rm ref}_c = E^{\rm ref}_{\rm tot} - E^{\rm EXX}_{\rm tot},
\ee
where $E^{\rm ref}_{\rm tot}$ is the exact total energy
obtained, e.g., from, an analytic, accurate configuration-interaction 
(CI) or quantum Monte Carlo (QMC) calculation, and EXX refers 
to the exact exchange. 
Here we have calculated the EXX energies in the
Krieger-Li-Iafrate~\cite{KLI} (KLI) approach~\cite{KLI} 
in the {\tt octopus} DFT code.~\cite{octopus}
The self-consistent EXX result -- the x-hole potential,
(spin) density, kinetic-energy density, and current density --
is used as input for our correlation functional.

Table~\ref{table1} 
\begin{table}
  \caption{\label{table1} 
Total energies from the full 
configuration-interaction calculations
(Ref.~\onlinecite{rontani}) for totally spin-polarized ($S_z=N/2$) quantum dots, 
exact-exchange total 
energies, the reference correlation energies 
($E_{\rm c}^{\rm ref}=E_{\rm tot}-E^{\rm EXX}_{\rm tot}$), 
$c_{\sigma\sigma}$ yielding $E_c^{\rm model}=E_{\rm c}^{\rm ref}$,
$E_c^{\rm model}$ obtained with a fixed average value $c_{\sigma\sigma}=1.32$,
and the LSDA correlation energy.
}
  \begin{tabular*}{\columnwidth}{@{\extracolsep{\fill}} c c c c c c c c}
  \hline
  \hline
  $N$ & $\omega$ & $E_{\rm tot}^{\rm ref}$ & $E^{\rm EXX}_{\rm tot}$ & 
  $E_{\rm c}^{\rm ref}$ & $c_{\sigma\sigma}$ & $E_c^{\rm model}$ & $E_c^{\rm LSDA}$ \\
  \hline
  3  & 1/4  & 2.081  & 2.103  & -0.0226 & 1.27 & -0.0245 & -0.0538 \\
  3  & 1/16 & 0.6908 & 0.7075 & -0.0167 & 1.41 & -0.0144 & -0.0382 \\
  6  & 1/4  & 7.233  & 7.296  & -0.0640 & 1.24 & -0.0750 & -0.1125 \\
  6  & 1/16 & 2.553  & 2.599  & -0.0458 & 1.36 & -0.0441 & -0.0795 \\
    \hline
  \hline
  \end{tabular*}
  \end{table}
shows the results for the same-spin case. Now the QDs are
completely spin-polarized with $S_z=N/2$. 
For each QD we show that value of $c_{\sigma\sigma}$ which yields the
reference correlation energy, i.e., $E_c^{\rm model}=E_{\rm c}^{\rm ref}$.
In these examples we find $c_{\sigma\sigma}=1.24\ldots 1.41$.
Thus, the variation of $c_{\sigma\sigma}$ is rather small in view
of the fact that the density parameter, defined in harmonic QDs 
as $r_s=N^{-1/6}\omega^{-2/3}$ (Ref.~\onlinecite{koskinen}), 
varies from $1.9$ to $5.3$. The second last column of
Table~\ref{table1} shows the correlation energy obtained
by using a fixed average value of $c_{\sigma\sigma}=1.32$. 
This leads to the maximum deviation of $\sim 23\,\%$ from 
$E^{\rm ref}_c$.
In comparison, the self-consistent LSDA correlation energy (last column)
deviates from the reference result by up to $130\,\%$. 

Table~\ref{table2} 
\begin{table}
  \caption{\label{table2} Similar to Table~\ref{table1} but for
unpolarized ($S_z=0$) quantum dots. The correlation energies
from our functional, $E_{\rm c}^{\rm model}$, have been calculated using 
the fixed average values $c_{\sigma\sigma}=1.32$ and $c_{\sigma\bar{\sigma}}=0.75$.
}
  \begin{tabular*}{\columnwidth}{@{\extracolsep{\fill}} c c c c c c c c}
  \hline
  \hline
  $N$ & $\omega$ & $E_{\rm tot}^{\rm ref}$ & $E^{\rm EXX}_{\rm tot}$ & 
  $E_{\rm c}^{\rm ref}$ & $c_{\sigma\bar{\sigma}}$ & $E_c^{\rm model}$ & $E_c^{\rm LSDA}$\\
  \hline
  2  & 1             & $3^*$          & 3.162  & -0.162  & 0.72 & -0.171 & -0.199 \\
  2  & 1/4  & $0.9324^\dagger$        & 1.046  & -0.114  & 0.82 & -0.102 & -0.139 \\
  2  & 1/16  & $0.3031^\dagger$       & 0.373 & -0.070 & 0.96 & -0.053 & -0.085 \\
  6  & 1/4  &  $6.995^\dagger$        & 7.391  & -0.396  & 0.73 & -0.406 & -0.457 \\
 12  & $1/1.89^2$ & $25.636^\ddagger$ & 26.553 & -0.917  & 0.71 & -0.983 & -1.000 \\ 
  \hline
  \hline
  \end{tabular*}
 \begin{flushleft}
  $^*$         Analytic solution by Taut from Ref.~\onlinecite{taut}.\\
  $^\dagger$   CI data from Ref.~\onlinecite{rontani}.\\
$^\ddagger$  Diffusion QMC data from Ref.~\onlinecite{pederiva}.
  \end{flushleft} 
 \end{table}
shows the results for a set of unpolarized $(S_z=0)$ QDs
in the range $1.5<r_s<5.7$.
Note that
for $N>2$ both same- and opposite-spin components of the correlation are present, 
and we have fixed $c_{\sigma\sigma}=1.32$ according to the conclusions above.
Fixing $c_{\sigma\bar{\sigma}}=0.75$ yields deviations of only $\lesssim 10\,\%$
from $E_{\rm c}^{\rm ref}$, except for the highly correlated case of 
$N=2$ and $\omega=1/16$, which shows a deviation of $25\,\%$. The LSDA 
is still considerably further off the reference result than our functional,
but it performs relatively better than in the polarized case discussed above. 
In particular, when $N=12$ the error in the LSDA correlation is only 
about $9\,\%$. This is in line
with the well-known fact that both the L(S)DA exchange and correlation,
respectively, become more accurate with increasing particle number.

\begin{figure}
\includegraphics[width=0.80\columnwidth]{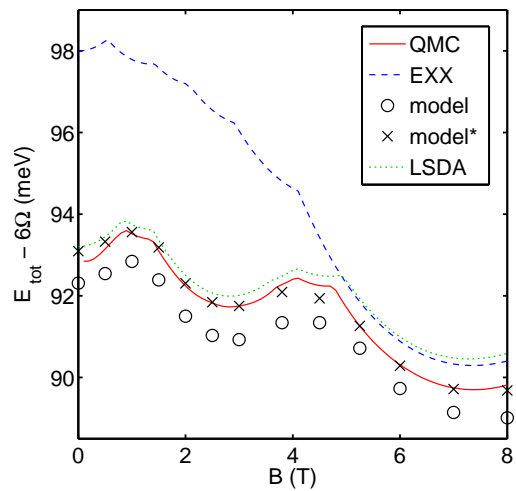}
\caption{(color online). Total energies (minus the confinement
energy) for a six-electron quantum dot as a function of
the magnetic field. Results are shown for the reference
quantum Monte Carlo calculations from Ref.~\onlinecite{henri}
(red solid line), for the exact-exchange
data (blue dashed line),
for our functional with 
$c_{\sigma\sigma}=1.32$ and $c_{\sigma\bar{\sigma}}=0.75$
(circles), for our functional with 
$c_{\sigma\sigma}=1.1$ and $c_{\sigma\bar{\sigma}}=0.7$ (crosses),
and for the local-spin-density approximation 
(green dotted line). The total energy is calculated
from our functional as 
$E_{\rm tot}^{\rm model}=E^{\rm EXX}_{\rm tot}+E_c^{\rm model}$.}
\label{fig1}
\end{figure}

In Fig.~\ref{fig1} 
we consider the total energy of a more general case: A six-electron QD as
a function of the magnetic field $B$ directed perpendicular to the dot plane.
 Increasing the field leads to non-trivial changes in the ground-state 
quantum numbers ($S_z,L_z$) and hence to ``kinks'' in the ground-state total
energy as a function of $B$. As the reference data, we use here the variational 
QMC results (red solid line) given in  Refs.~\onlinecite{henri} 
and \onlinecite{nicole} for a wide 
range of $B$ up to total spin polarization. 
The confinement strength is here
$\omega=0.42168$, corresponding
to a typical confinement of 5 meV when modeling QDs in GaAs.~\cite{qd}
Note that the {\em total} confinement energy, i.e., 
$6\Omega = 6\sqrt{\omega^2+\omega_c^2/4}$, where $\omega_c=B/c$, has been 
subtracted from the total energies to clarify the comparison. 
We point out that the variational QMC method gives an upper bound
for the true total energy. On the basis of previous comparisons
between the variational and diffusion QMC, and exact diagonalization,~\cite{ari_wigner}
our reference data in Fig.~\ref{fig1} can be expected to overestimate 
the exact total energy by {\em at most} $0.2\ldots 0.3$ meV. The maximum
possible errors are smaller in the polarized regime ($B\gtrsim 5$) T.

Overall, Fig.~\ref{fig1} shows reasonable agreement of our functional
(circles) with the QMC data through the full range of the magnetic field. 
However, the functional yields systematically slightly too low correlation 
energies, and thus too low total energies, even if the possible
overestimation of the total energy given by the variational QMC is 
taken into account. We point out that 
the functional is here applied with the fixed parameters
$c_{\sigma\sigma}=1.32$ and $c_{\sigma\bar{\sigma}}=0.75$ 
suggested by the results in Tables~\ref{table1} and \ref{table2},
which correspond to considerably weaker confining potentials 
(smaller values of $\omega$).
Obviously, this difference implies 
a different average electron density, and thus a different range 
of the relative correlation energy. In fact, 
if the parameter values are reduced to 
$c_{\sigma\sigma}=1.1$ and $c_{\sigma\bar{\sigma}}=0.7$,
excellent agreement with QMC is found
(see the crosses in Fig.~\ref{fig1}).
Hence, it seems that particularly high precision of our functional would require
modeling of $c_{\sigma\sigma}$ and $c_{\sigma\bar{\sigma}}$ as a function of
the particle density. Most importantly, however, 
the systematic performance of our approximation(s) in Fig.~\ref{fig1} demonstrates
that the magnetic-field effects, electron currents, and spin are correctly
accounted for. 

\begin{figure}[th!]
\includegraphics[width=0.80\columnwidth]{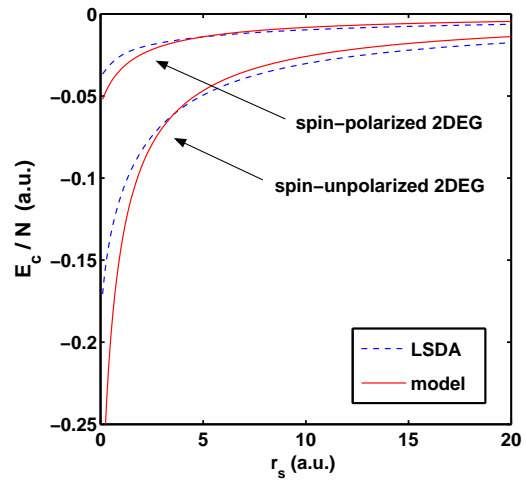}
\caption{(color online). Correlation energy per electron
in a homogeneous two-dimensional electron gas of full
spin polarization (upper curves) and zero polarization 
(lower curves). The solid lines show the result from our functional
with the original average parameter values $c_{\sigma\sigma}=1.32$ and
$c_{\sigma\bar{\sigma}}=0.75$. The dashed lines show
to the local spin-density approximation for the correlation, 
corresponding to the numerically exact result in this 
system.~\cite{attaccalite}
}
\label{fig2}
\end{figure}

We note that the good accuracy of the 
LSDA in terms of total energies (see the dotted line in Fig.~\ref{fig1}) 
is due to the compensation of respective 
errors in the exchange and correlation energies.~\cite{nicole}
On the other hand, we tested our correlation functional
for the important limit of the {\em homogeneous} 2D electron gas
(2DEG), for which the LSDA correlation is exact, and found
reasonable agreement as a function of $r_s$ 
for both zero and full spin polarization (see Fig.~\ref{fig2}).
Here we used the original average parameter values $c_{\sigma\sigma}=1.32$ and
$c_{\sigma\bar{\sigma}}=0.75$. Note that according to the numerically
exact results in Ref.~\onlinecite{attaccalite}, the ground-state of the 2DEG 
is unpolarized for $0 < r_s \lesssim 26$.

Finally, we point out, 
that in principle a given functional should be evaluated with
KS orbitals obtained from self-consistent calculation instead of a post-hoc
manner as we have done in this work. However, the variational nature
of DFT implies 
that if one evaluates the total energy with densities which slightly differ
from the self-consistent one, the resulting change in the energy
is of second order in the deviation of the densities.

\section{Conclusions}

We have derived a spin- and current-dependent
approximation for the correlation energy of finite 
two-dimensional electron systems. The core of the derivation 
is a model for the correlation-hole function of both
same-spin and opposite spin pairs, respectively, that
satisfies a set of exact properties. The excellent results obtained for 
few-electron quantum dots with different spin-polarization, current, 
external magnetic field, and covering a wide range of correlation energies,
strongly recommend further developments along the 
construction we have presented here.

\vspace{5mm}

\begin{acknowledgments}
We thank Ari Harju for the variational quantum Monte Carlo data.
This work was supported by the Deutsche 
Forschungsgemeinschaft, the EU's Sixth Framework
Programme through the Nanoquanta Network of
Excellence (NMP4-CT-2004-500198), and by the Academy of
Finland. C. R. P. was supported by the European
Community through a Marie Curie IIF (MIF1-CT-2006-040222) 
and CONICET of Argentina through PIP 5254.
\end{acknowledgments}


\begin{thebibliography}{ll}
\bibitem{units}
We use the effective atomic units (a.u.) with the
material parameters for GaAs: $m^*=0.067\,m_e$ and $\epsilon=12.4\,\epsilon_0$.
\end{thebibliography}
\end{document}